\documentclass[aps,amsmath,twocolumn,nofootinbib]{revtex4-2}
\usepackage{graphicx}
\usepackage{relsize}
\usepackage{enumitem}
\usepackage[table,usenames,dvipsnames]{xcolor}
\usepackage{ifthen}
\definecolor{linkcolor}{rgb}{0.6,0,0}
\definecolor{citecolor}{rgb}{0,0,0.75}
\definecolor{urlcolor}{rgb}{0.12,0.46,0.7}
\usepackage[breaklinks, colorlinks=true, urlcolor=urlcolor, linkcolor=linkcolor,citecolor=citecolor,pdfencoding=auto]{hyperref}
\hypersetup{linktocpage}
\setlist{nolistsep,leftmargin=*} 

\def\fraction#1/#2{\leavevmode\kern.1em
 \raise.5ex\hbox{\the\scriptfont0 #1}\kern-.1em
 /\kern-.15em\lower.25ex\hbox{\the\scriptfont0 #2}}

\makeatletter
\renewcommand\@makefnmark{\hbox{\@textsuperscript{\normalfont\color{linkcolor}\@thefnmark}}}
\makeatother

\begin{document}


\title{The AI crisis for teaching:\\
``Train your \textit{own} neural network!''}

\author{Douglas Scott} \email{dscott@phas.ubc.ca, and not afrolop@phas.ubc.ca}
\affiliation{Dept.\ of Physics \& Astronomy,
 University of British Columbia, Vancouver, Canada}

\date{1st October 2026}

\begin{abstract}
Everyone seems to be thinking about so-called artificial intelligence (AI)
these days.  As scientists, we understand that we're not about to have a
conscious computer intelligence taking over the world, but we're nevertheless
concerned about how ``large language models'' (LLMs) and AI ``agents'' are
impacting the way that we do science.
This concern has generated several recent essays about
the effects of AI on research, including in my own field of astronomy and
astrophysics.  These opinion pieces
tend to miss the biggest current concern for those whose
jobs combine teaching with research -- although the threat to future
scientific research as we know it is probably a bit concerning,
the effect on teaching is an existential crisis right now!
We need to hang on to the parts of education that we think are at its core --
so that students coming out of degree programmes are actually able
to think for themselves, ``like an astrophysicist''.  This means emphasising
to students that they should use AI tools to {\it support\/}
their education and not to {\it replace\/} it.
One way of explaining the importance of this principle is to point out
that students should be training their own neural networks.
\end{abstract}

\maketitle

\section{The problem\texorpdfstring{$\color{linkcolor}^0$}{}}
\footnotetext[0]{Because this is not a 1st April paper, I will try to
keep a serious tone in the main text and leave
flippant comments to these footnotes.}
Machine learning, neural networks, and so-called ``artificial intelligence''
(AI)
are changing lots of things. Many scientists have written about the effects
of ``large-language models'' (LLM) and AI ``agents'' on the way that science is
carried out, with speculation about how it will be done in the future.
To focus on my field, astrophysics, several recent essays have
discussed the practical and sociological repercussions of these new tools for
research \cite{LLMsResearch}.

I concentrate here on astrophysics because it is what I study and mostly what
I teach.  The issues will be quite similar in other sub-fields of physics, and
broadly similar in other sciences.  On the other hand, the situation
might be very different in other academic disciplines.\footnote{Let's start
with a Douglas Adams reference: this situation is what he
describes as ``Someone Else's Problem'', which is cloaked by the SEP field,
and hence not seen \cite{adams}.}

The norms for doing astrophysical research are indeed changing, and all
research scientists are currently grappling with how to cope with
that.\footnote{Ironically, organisers of some AI-based journals and
conferences have been complaining the most loudly.  Those of us in other
science fields find it difficult to be sympathetic.}
However, I think this discourse misses the more important issue for
university-based
academics who teach as part of their jobs: there certainly are thorny issues
related to research, but these are dwarfed by the problems involving teaching,
which is now in a genuine crisis.  Perhaps even an existential one.

We cannot teach like we used to, especially when it comes to student
assessment.  We need to be clear about the value gained through the
experience of having a real human being teaching course material.  We also
need to find ways of certifying that students leaving our courses
genuinely know something.  If we don't deal with these issues, we may not be
teaching at all in the future.

\section{The new normal}

Here is an example of the problem.
I recently taught at an after-school programme for high-school students who
are enthusiastic about physics.  I gave them exercises involving working out
very large numbers that might come up in cosmology: calculating the
size of the observable Universe, then moving onto estimating how big a number
might be required to describe the positions of all the atoms over all of
history.  Students sitting just a few feet away from me simply typed the
question into a device and read out the answer --\footnote{No em-dashes in this
article.} with no attempt to hide the
fact that this is how they answered my question.  They clearly thought that
this was what I was asking them to do!

Let me give another example, from the early days of LLMs.  In an upper-level
astrophysics course, I assigned a homework problem where students were asked
to perform a calculation whose answer is well known to be
\fraction1/2.\footnote{In
case anyone is interested, it was the calculation of what fraction of the
distance to a source has the maximum probability for gravitational lensing.}
A student posted a question on the course Q\&A website, stating that they
tried several times and kept getting \fraction2/3 rather than \fraction1/2,
so what were they doing wrong?  I checked myself and confirmed that it is
\fraction1/2, like it had always been in the past.  But then I asked ChatGPT
the same question and it confidently said ``\fraction2/3'',
giving a line-by-line calculation with at least one
algebraic error.  The student had clearly just asked ChatGPT and then
put more effort into querying the correct
solution than actually doing the work -- and no learning was happening.

Talking more generally, we all know that it can be great to have an
LLM answer complex questions for you; hence, there are obviously aspects of
AI that could be helpful for students in our university courses.
However, the overwhelming view of instructors is that students are not using
these tools to improve their learning, but are avoiding mental effort
by using AI to produce homework answers that are not their own and do not even
involve ideas passing through their brains.\footnote{Or in other words
``cheating''.}

There is a basic approach to teaching
that has been effectively used in universities
for centuries to teach students in the physical sciences\footnote{I would not
claim that everyone teaches in the same way, but the idea is pretty universal
that physics students need to go off and struggle in order to understand
difficult ideas.} -- the didactic
method coupled with self-study.  That system has been broken by the
availability of LLM-generated answers.  University lecturers now regularly
complain about how AI is being used for assignments, which can therefore
no longer be trusted to function as
practice exercises for students to learn how to treat new concepts.

One response to this change is to let the students sink themselves if they
want.  Some ``old-school'' colleagues
say things like ``this year, the average on my
homework was 98\%, but the average on the final exam was 37\% --
so I just failed more people''.

On the other hand, some colleagues who see how easy it is to find out stuff
using LLMs are telling students ``just go for
it'', encouraging them to use whatever tools they can in order to find
solutions to problems.  I think this confuses the situation: it is different
for an expert to ask an LLM to answer something in detail when they already
know a lot about the topic, compared to a student just starting out on their
educational journey, who knows very little, and thus is incapable of assessing
whether the LLM answer is reasonable.\footnote{As an example of how
so-called ``AI'' isn't really ``intelligence'', in the early days of LLMs, I was
teaching a course where the previous instructor's notes referred to an ``RCL''
circuit, which I felt sure I'd always called an ``RLC'' circuit.  My question
to the search engine was forwarded to some bot, which answered ``An RCL circuit
is different from an RLC circuit: an RCL circuit consists of a resistor, a
capacitor and an inductor, whereas an
RLC circuit consists of a resistor, an inductor and a capacitor''.}

For a long time, instructors have been encouraged to move away from formal
exams and towards a system with more continuous assessment.  However, it now
seems impossible to give course assignments and have any confidence that the
students are being tested in any way, other than perhaps
how well they can cut and paste into a box.  The concern is that we will
end up just teaching students how to ask an LLM the right question, in other
words we will focus on ``prompt engineering'' rather than learning and
understanding.

I have taught courses for more than 30 years, and so perhaps
I should be coasting by now.  However, the best instructors I know
are always asking themselves if their teaching is effective, and hence their
approach is always evolving.  This surely has to be even more the case now --
we all have to learn how to adapt to the challenges posed by AI in
the classroom.

What will astrophysics courses look like 5, 10 or 20 years from now?  If anyone
at all
can appear to get expert answers to difficult questions without actually
knowing anything, is there any point to taking a university course?  Will
students see any value in enrolling in our course offerings,
or indeed university degrees
as a whole -- especially considering how expensive that can be?
Is it possible that campuses will be empty by the middle of this century?

\section{The optimistic view}
Of course not everything about AI is bad.  Some instructors are trying to
focus on the positive aspects and figure out how to use AI in the classroom
to enhance learning rather than replace it.  However, doing this effectively is
challenging.\footnote{I remember in the early days of LLMs, there was a
discussion among faculty members at my own institution and one Arts professor
said ``This is great for teaching students how to write essays -- when I give
them an essay prompt now, I get them to {\it either\/} use an LLM to
make a list of bullet points from the prompt, {\it or\/} they can get an LLM
to generate text from those bullet points -- and I insist that they
promise not to use it for both!''}

Undoubtedly there are tasks for which AI can be very useful.  If you know a lot
about a topic, you can ask a detailed question of an LLM and use your
expertise to decide how much you believe the answer, perhaps chasing down
original references to ensure that it's all correct.
At a more basic level,
using a souped-up grammar checker is surely fine -- particular if you have to
write in English and it is not your first language.\footnote{My own first
language is Scots.}
Asking an LLM to reduce your abstract by five words also seems like a
reasonable use.  Getting the AI-assisted
search engine to look through Stack Exchange for an answer to your computing
question is clearly more efficient than either searching it yourself or
posting the question and waiting for a human to
answer.\footnote{Although the number of
new posts to Stack Exchange has plummeted recently -- so there are no new
answers, just recycled LLM concoctions.}
But where do you draw the line?

Another argument made in favour of accepting the evolving world is
that this AI revolution is just the latest in a series of changes in how we get
information.  Many of the major upheavals in the past
brought out the doomsday predictors, but turned out not to cause the end
of civilisation as we know it.  For example, there were probably naysayers
when counting was first invented,\footnote{``You should just picture a lot of
sheep, rather than using this new-fangled business of putting scratches on a
stick''.}
and certainly
some people claimed that the sky was falling when pocket calculators replaced
mental arithmetic.  Additionally, the availability of information on the
internet has made it usually unnecessary to go to a library, and yet we've
adapted to that; so mostly we see the advantages rather than worrying that
the change is catastrophic.  Is the AI revolution any different?  I think it is.
For one thing, we are now in a position to ask a question and get an
extremely ``confident'' and completely wrong answer.  That wasn't true for
counting, the calculator,\footnote{OK, you could type the wrong numbers into a
calculator, but that's not the same thing.  If you give a calculator the
right input, you'll get the right answer.  And if you dispute that answer, your
calculator won't say ``you're absolutely correct to query 1+1=2, and I now
see that this was a hallucination, since the correct answer, as you point out,
is really 3''.} the internet, or any previous
developments in how we deal with information.

\section{What do we think we're doing in the classroom anyway?}

Given that things are changing whether we like it or not, we need to think
carefully about the point of the whole teaching endeavour -- and then we need
to try to preserve the core aspects.  So, what are we actually trying to do
when we stand in front of a university class?

In the physical sciences, we tend to introduce students to a concept,
explain it in some different ways and then give several practice problems using
that concept.  It is through this practical application of each new idea
to problems that we consider the concept has been fully understood.
The idea needs to be worked through in the brain of each student, taking both
time and effort, in order for them to ``get it''.

The situation may be quite
different in other disciplines, even within the sciences.  Perhaps it's
possible to learn the main goals of some non-physics subject by fully
embracing AI tools, but that is not the case in astrophysics, where
getting an LLM to do all the work does not meet the criteria that instructors
have for what ``learning'' means.\footnote{In discussions at my university,
I've seen instructors in other fields reworking their courses to encourage
students to use AI for absolutely everything, in a way that frankly horrifies
me.}

We teach courses and assign passes and numerical grades to indicate how well
a student did on coursework and exams.  Presumably, passing a course should
certify that a student has a particular level of knowledge and proficiency
in that topic.  The numerical grade gives information about how well
the student fulfilled the course goals -- describing certain
concepts and being able to carry out specific kinds of calculation.
At the end of a degree programme, a successful student gets a diploma,
which surely means we think they are
qualified in some way in this subject area.  It does not mean that they simply
know to ask an LLM how to answer questions on this topic.

Perhaps there has been a fault in our system because we haven't been giving
grades for what really matters?  Now we need to focus on evaluating
students based on the main goals of a course.  If high grades are assigned for
getting the right answer in the shortest time, then we shouldn't be surprised
if students use an LLM on their homework.  Instead we should think about how
to reward actual learning.\footnote{Some students who saw a draft of this
article pointed out that some of them do in fact want to learn, and would
prefer that instructors {\it discourage\/} AI use, so that they can actually
do stuff themselves -- we should clearly be doing whatever we can to encourage
such students.}

Additionally, there is a notion in astrophysics that we are teaching
students a certain suite of problem-solving skills that can be applied to
a wide range of different situations.  We might call this ``thinking like an
astrophysicist''.\footnote{Perhaps a bit like what physicists call doing
``Fermi problems'', but applied to solving problems using multifarious
approaches.}
Even if it's not simple to explain exactly
what it means, there's broad consensus on the ideal that we're aiming
for: we would like students to be able to approach the sorts of thinking that
the best astrophysicists make look easy (we all have our own examples of these
people\footnote{I think of the most inspiring scientists
from my early career.}).
This approach to treating concepts and data has meant that students graduating
with degrees in astrophysics readily find employment in many different jobs
where problem-solving skills are valued.

\section{Where do we go from here?}
One reaction to the current crisis would be to concede, accepting
that students will get answers to all their questions from a computer, with
very little in the way of brain
activity,\footnote{We've all read of studies showing that brain activity when
solving  problem with an LLM is close to zero compared with trying to think it
through for yourself (I was going to look up these studies, but couldn't figure
out how to do it without using AI!)}
and that's just the way it is now.
To go further we might give up trying to actively teach students anything in
future.\footnote{A few years ago, an English professor wrote in an opinion
piece that he no longer felt capable of doing the job that he had been hired
to do, namely to teach students how to
interpret text, form coherent opinions and write well-formed and well-reasoned
discussions. Because of this he was simply taking early retirement.}

Encouraging students to put in the effort for themselves is not a
new problem, of course.
I used to refuse to give out solutions to problems (for example, a practice
exam) until students had tried to find solutions for themselves.  If I posted
model answers right away, some students simply looked at those solutions
and convinced themselves that they could do the problem.\footnote{Perhaps by
pressing the solutions to their foreheads.}
However, when faced with
a blank piece of paper in an exam, they were completely lost.
Similarly, I used to be concerned about finding examples that were not
already solved on some
internet site.\footnote{One of the positive consequences of the rise of AI in
universities is that sites set up to help students cheat have essentially
disappeared.  For legal reasons I don't want to name names, but there was one
that started with the same letters as ``chicken'' and rhymed with ``egg''.}
So what's the
situation now?  It seems clear to me that it's even worse!  We somehow have
to convince students to solve things for themselves, although almost
instantaneous solutions to virtually every problem are now easily available.

What can we do to respond to this new reality?
In my department, instructors are trying different things, with a few
common strategies.
\vskip 5pt
\begin{enumerate}[start=0]
\item Continue as before.
\item Stop grading homework.
\item Put more weight on formal exams.
\item Use device-free quizzes.
\item Give oral exams.
\end{enumerate}
\vskip 5pt
The first suggestion seems completely hopeless and leads to
meaningless grades (or worse, grades that reflect how much money a student has
spent on AI access) and university degrees that aren't worth the paper they're
printed on.  Suggestions 2 and 3 are what I've done in my courses,
basically because I don't feel like I've got much choice.  Homework exercises
are now for participation marks only, and students are reminded regularly
that if they don't use the homework to practice how to do problems,
they'll perform extremely poorly on the exams, which now have to be worth a
larger fraction of the total grade.  However, even in formal exams, the
surreptitious use of AI is an increasing
problem.\footnote{A colleague recently reported that they had seen their first
case of ``smart glasses'' used in an exam.  There are also devices called
things like ``magic calculators'', which can look like simple calculators,
but are actually phones that can take a photo of the exam question and get
AI to give the answer.}

Traditionally, teaching assistants (TAs) spent a lot of their time grading
homework, lab reports, etc.  They would typically write a few words in red pen
on each student's work, to explain where they went wrong.  These days,
a lot of students are using AI to do the assignments, despite efforts to
convince them that it is ultimately counter-productive to do so; hence,
taking time to go through the submitted work is becoming pointless.  On the
other hand, students can get quite detailed feedback using an AI tutor --
better than the brief comments from the TA.  The abandonment of grading
homework means that TA time can be freed up to provide more tutorials where
they interact directly with the students; so that is potentially a positive
outcome for AI-driven changes to the system.

Let me say a little more about item 3 above.
One solution to the ``homework'' problem is to set up a system where students
have to answer questions without any computational aids.  These have
various names, but at my institution they are
called computer-based test facilities (CBTFs) or computerised examination
centres (CECs).  This is typically a room with dumb computers, which are
connected only to a testing system.  Students taking quizzes on these computers
have to leave their
devices at the door.  If things are set up well, then it's possible for
students to do their assignments at a range of times, rather than during
scheduled course time (which otherwise gets filled up with in-class
quizzes).  Hence,
it acts as a way of providing continuous assessment of student understanding
without any AI use.  How easy it will be to police this remains to
be seen, as devices get smaller and
smarter.\footnote{In a faculty discussion at my university, there was a joking
suggestion that, in addition to having metal detectors for exams, we should
make students come in swimsuits and do a length of the pool on the way into
the exam room!}

Lastly, there is item 4 in the above list, giving oral exams.  This feels like
going back to the Socratic method.  It can be effective, but only if enough
time is scheduled to give each student a fair chance of shining.  It also runs
counter to the reduction in student anxiety that we also have to consider.
Moreover, it's obviously a challenge to scale this to a large class, since
if different students are asked different questions in a short interview, then
it's hard to avoid claims that it was unfair.\footnote{And asking the same
questions over and over again is really boring for the instructor.}

These are some strategies for responding to the evaluation crisis, and there
are surely more.\footnote{Things are developing so rapidly that the list that
I gave is probably already out of date.}
The crucial thing is to think carefully about what we are doing when we teach
an astrophysics class -- to figure out the essential ingredients -- and
then do everything we can to hold onto those things!  We need to continue to
get students to use their own brains to develop problem-solving skills,
inspired by some of the most exciting topics in science (extreme physics
conditions, black holes, dark matter, the possibility of life on other
planets, etc.), leading to capabilities that make our graduates appealing to
employers.

\section{Final thoughts}

Let me end by getting a bit philosophical.  Is it OK to end up with a
world where nobody thinks for themselves
any more?\footnote{For a prophetic example of early concerns over computers
replacing human brain power, see Isaac Asimov's short story,
``Feeling of Power'' \cite{asimov}}
I'm pretty sure that most of us don't think so.  The reason is the same
for scientists as for people in general, namely, that interactions with
other humans are a crucial part of what makes us what we are.  Because of this,
I'd like to believe that being in the class with me explaining stuff is more
valuable than just reading
a textbook.\footnote{Or not even reading the textbook, but getting an
AI-generated summary.  And then not even reading that, but getting the
computer's voice to read it out for you!}

What does it mean to be human?  
I'm no longer as young as I used to be,
and when I reflect on my time in teaching, and life in
general, it seems that directly connecting with other human beings is
central to all of that.
For me, connecting involves telling stories in one
way or another,\footnote{In my case, funny stories are what make it fun for
me to teach, and my attitude is that if I'm not enjoying teaching, then the
students are almost certainly not enjoying it either.}
to explain concepts, contexts and how to think
about them.\footnote{I'm reminded of the aliens in Star Trek called the
Tamarians, who communicate using allegories, like ``Temba, his arms wide'' --
the idea of the episode being that the Federation universal translator was
unable to deal with this way of conveying meaning and the humans had to learn
for themselves how to talk with the Tamarians.}
Let's not give up on the experience of direct face-to-face teaching.

The most diligent students have {\it not\/} been using AI to do all their
thinking for them -- they actually want to learn!  We need to get the typical
student to understand the value of this approach to taking our courses.
And that means encouraging good uses of AI to advance {\it human\/}
learning.\footnote{See the article ``Deeper Learning in Astronomy'' by
Scott \& Frolop \cite{HL}}

Smart students can use artificial ``tutors'' to manage their learning,
drip-feeding them course material, customising quizzes and giving feedback
tailored to the student's level in the learning process.  Some course
instructors have developed ``chat-bots'' that are trained specifically on the
course materials and will allow students to discuss the topics in detail,
perhaps
even monitored by the instructors or TAs.\footnote{Setting up a bespoke
chat-bot is a lot of work for course instructors, and hence isn't a
labour-saving solution.  On the other hand, maybe that suggests that
it's worth doing!}  Many products are
available for students to effectively have their own study buddy, and these
will only get better.  If they help get the right ways of thinking into
students' heads faster, then they're obviously a good idea.

We certainly want to make sure that, when a student passes a course or gets a
degree certificate, the stamp in the corner, the imprimatur of the
university, actually means something.  And that something is that the
student has achieved a set of learning goals and could in principle work
their way through certain kinds of problems even on a desert island with
nothing but their own wits available.

The basic AI policy for my classes is
``feel free to use AI tools as aids to your learning, but {\it not\/} as
a replacement for learning -- and {\it you\/} are responsible for this!''
As a short instruction for students, I've been been repeating the mantra that
the point of students being in my course in a university degree programme is
to actually learn stuff for themselves.  The slogan is:
``Train your own neural network!''

\vspace{1.2cm}
\noindent{\bf Acknowledgements}:\ 
I would like to thank many colleagues for interesting discussions on this
topic.\footnote{And in particular for sharing amusing stories about the uses
and abuses of AI.}
No AI was used to create this article.

\end{document}